\documentclass[aps,prb,showpacs,raggedbottom,nobalancelastpage,amssymb,twocolumn,groupedaddress]{revtex4}
\usepackage{graphicx}
\usepackage{amsmath}
\usepackage{amsfonts}
\usepackage{amssymb}
\usepackage{epsfig}
\usepackage{color}
\usepackage{dsfont}

\newcommand{\abs}[1]{\left\vert#1\right\vert}

\newcommand{\bra}[1]{\langle#1\vert}
\newcommand{\ket}[1]{\vert#1\rangle}

\begin{document}

\title{Dynamical Casimir Effect and Work in Fermionic Fields}

\author{Gianluca~Francica}
\email{gianluca.francica@gmail.com}
\noaffiliation

\date{\today}

\begin{abstract}
We consider a quantum massless fermionic field in (1+1) dimensions in the case of moving boundaries. We work in the canonical approach in order to find a Hamiltonian describing the dynamics of the field. Thus, we study the statistics of work and particles produced from the vacuum by a linear driving of a boundary.
\end{abstract}

\maketitle

\section{\label{sec.intro}Introduction}
The dynamical Casimir effect is the phenomenon for which moving boundaries of a quantum field could induce particle creation from the vacuum. It was originally foreseen by Moore, which has used a simplified one-dimensional model~\cite{moore70}, and afterwards studied by DeWitt\cite{dewitt75}, Fulling and Davies~\cite{fulling76,davies77}, and many others.
Anyway, differently from the bosonic fields for which there have been many studies (see, e.g., Ref.~\cite{dodonov20} for a recent review), the  dynamical Casimir effect in fermionic fields remains less investigated~\cite{mazzitelli87}.

An interesting question is how the dynamical Casimir effect is related to the thermodynamics of the process. Recently, many studies has consolidated the emerging field of quantum thermodynamics~\cite{kosloff13,Vinjanampathy16,bookthermo18}.  In this perspective, the statistics of the work performed and the change of photon number has been studied in the case of a cavity with an oscillating boundary~\cite{fei19}.

In this paper, we aim to study the dynamical Casimir effect in the less investigated case of a massless fermionic field in $(1+1)$ dimensions, by adopting a canonical approach previously employed for the bosonic case~\cite{schutzhold98}. In particular, we study the statistics of work and particles produced from the vacuum by a linear driving of a boundary.

\section{\label{sec.model} Model}
We consider a quantum fermionic field in $(1+1)$ dimensions between moving boundaries. In the massless case the Dirac Lagrangian density being
\begin{equation}
\mathcal L = \frac{i v}{2} \left(\bar{\psi}\gamma^\alpha\partial_\alpha\psi-(\partial_\alpha\bar{\psi})\gamma^\alpha\psi\right)
\end{equation}
where $\partial_\alpha=\partial/\partial x^\alpha$ with $x^0=v t$ and $x^1=x$. The adjoint field $\bar{\psi}$ is defined by $\bar{\psi}=\psi^\dagger \gamma^0$, and the Dirac matrices satisfy the anti-commutation relations $\{ \gamma^\mu,\gamma^\nu \}=2 g^{\mu \nu}$ with $g^{\mu \nu}=\text{diag}(+,-)$. The Dirac matrices can be taken to be two-dimensional, we consider $\gamma^0=\sigma_x$ and $\gamma^1=-i \sigma_y$ where $\sigma_\alpha$, with $\alpha=x,y,z$, are the Pauli matrices.
We assume that the boundaries are placed at the points $x=0$ and $x=l(t)$ and we consider the bag-like boundary conditions~\cite{mazzitelli87} 
\begin{equation}
e^{i\theta \gamma_5}\psi = i n_\mu \gamma^\mu \psi
\end{equation}
where $\gamma_5=\gamma^0\gamma^1=\sigma_z$ and $n^\mu$ is the unit, exterior normal to the boundary. 
In particular, we will put $\theta=\theta_0$ constant for $x=0$ and $x=l(t)$.

The energy is $E=\int T_{00} dx$ where $T_{00}$ is the density
\begin{equation}
T_{00}=-i v \psi^\dagger \gamma_5 \partial_x \psi
\end{equation}
The instantaneous eigenfunctions $\psi_n$ of the one-particle energy $-i v \gamma_5 \partial_x$ are such that $-i v \gamma_5 \partial_x \psi_n= \omega_n \psi_n$. We find
\begin{equation}
\psi_n = \frac{1}{\sqrt{2l}}\left(
                              \begin{array}{c}
                                e^{i\frac{\omega_n}{v}x} \\
                                i e^{-i\left(\frac{\omega_n}{v}x-\theta_0\right)} \\
                              \end{array}
                            \right)
\end{equation}
which are normalized such that $\int \psi_n^\dagger \psi_m dx = \delta_{m,n}$, with one-particle energies $\omega_n = (n+1/2)\pi v/l$.
In order to quantize the Dirac field we expand it in terms of the eigenfunctions $\psi_n$ as
\begin{equation}
\psi = \sum_{n\geq 0} a_n \psi_n + b_n^\dagger \psi_{-1-n} = \sum_n q_n \psi_n
\end{equation}
such that automatically satisfies the boundary conditions.
The Lagrangian $L=\int \mathcal L dx$ reads
\begin{equation}
L=\sum_n\left(\frac{i}{2}(q^\dagger_n \dot q_n-\dot q^\dagger_n q_n)- \omega_n q^\dagger_n q_n\right) + i \sum_{n m} q^\dagger_m M_{mn}q_n
\end{equation}
where we have defined the matrix elements
\begin{equation}
M_{mn}= \frac{1}{2}\int (\psi_m^\dagger \dot\psi_n -\dot\psi_m^\dagger \psi_n )dx
\end{equation}
By considering $q_n$ as independent variables, their canonical conjugate momenta will be $p_n=\partial L/\partial \dot q_n=i q_n^\dagger/2$. Thus the Hamiltonian $H$ describing the dynamics reads
\begin{equation}
H= E - i \sum_{n m} q_m^\dagger M_{mn}q_n
\end{equation}
where the energy is $E=\sum \omega_n q_n^\dagger q_n$ and $M$ needs to be anti-hermitian in order to have $H$ hermitian.
We impose the anti-commutation relations $\{a_n,a_m^\dagger\}=\{b_n,b_m^\dagger\}=\delta_{m,n}$ and $\{a_n,a_m\}=\{b_n,b_m\}=\{a_n,b_m\}=\{a_n,b_m^\dagger\}=0$. By subtracting the zero-point energy, we have $E=\sum \abs{\omega_n}(a_n^\dagger a_n+b_n^\dagger b_n)$ and the ground-state is the vacuum state $\ket{0}$ defined by $a_n\ket{0}=b_n\ket{0}=0$.

The elements of the matrix $M$ are
\begin{equation}
M_{mn}=\frac{(-1)^{m-n}}{2}\frac{m+n+1}{m-n}\frac{\dot l}{l}
\end{equation}
for $m\neq n$, and $M_{nn}=0$, such that $M^T=-M$.
We observe that, since $E\ket{0}=0$ at any time, only due to the presence of the matrix $M$ it is possible to acquire a dynamical Casimir effect.

\section{\label{sec.stat} Dynamics and Statistics}
We focus on a linear changing of the boundary $l(t)=l_0+\alpha t$, such that $[H(t),H(t')]=0$ and the time-evolution operator is $U=e^{-i \tilde H \Delta l}$, where $\Delta l = \ln (l(t)/l(0))$ and $\tilde H$ is time-independent and such that $H = \tilde H \alpha/l(t)$.
In order to study the dynamical Casimir effect, we consider the initial state $\ket{0}$. In this case the distribution of work performed is equal to $p(w) = \abs{\bra{n}U\ket{0}}^2 \delta(w-\epsilon_n(t)) $ where $\ket{n}$ is the eigenstate of $E(t)$ with eigenvalue $\epsilon_n(t)$. The characteristic function $\chi(u) = \int e^{i u w} p(w) dw$ can be expressed as
\begin{equation}\label{eq.chara}
\chi(u) = \bra{0} e^{i u \tilde E }\ket{0}
\end{equation}
where $\tilde E = U^\dagger E(t) U$ and the work moments are given by $\langle w^n\rangle =(-i)^n\chi^{(n)}(0)$. We note that the work is completely irreversible since the vacuum energy remains equal to zero. Similarly, the characteristic function of the particle created is $\chi_N(u)=\bra{0} e^{i u \tilde N }\ket{0}$ with $\tilde N = U^\dagger N U$ where $N$ is the number operator $N=\sum a^\dagger_n a_n + b^\dagger_n b_n$.

For small $\Delta l$, we get
\begin{equation}
U\ket{0} \approx \ket{0} +\frac{\Delta l}{2} \sum_{m\geq 0, n\geq 0}(-1)^{m+n}\frac{m-n}{m+n+1}a_m^\dagger b_n^\dagger\ket{0}
\end{equation}
such that couples of fermions $(a,b)$ are created from the vacuum. In particular, it is more likely  to create fermions $a$ with high energy and $b$ with low energy, and viceversa to create fermions $b$ with high energy and $a$ with low energy.

Lets show how to calculate the characteristic functions. We start by noting that the operator $\tilde H$ can be expressed as
\begin{equation}
\tilde H = \sum_{mn}\left(c^\dagger_m A_{mn} c_n + \frac{1}{2}(c^\dagger_m B_{mn}c_n^\dagger + h.c.)\right)
\end{equation}
where $c_n=a_n$ and $c_{-n-1}=b_n$ for $n\geq 0$, and the matrices $A$ and $B$ are such that $A^\dagger=A$ and $B^T=-B$. In particular $A_{nn}= \abs{n+1/2}\pi v/\alpha$ and for $n\neq m$, $A_{mn}=-i M_{mn} l/\dot l$ if $m,n\geq 0$ or $m,n<0$ and $A_{mn}=0$ otherwise, $B_{mn}=-B_{nm}=-i M_{mn} l/\dot l$ for $m\geq 0$ and $n<0$ and $B_{mn}=0$ otherwise.
In order to solve the dynamics, we note that the operator $\tilde H$ can be diagonalized by a Bogoliubov transformation~\cite{lieb61} $c_n = \sum_k u_{nk}\eta_k + v_{nk}\eta^\dagger_k$ such that $\tilde H = \sum \Lambda_k \eta_k^\dagger \eta_k$. The matrices $u$ and $v$ are defined by $A u+B v^*=\Lambda u$ and $Av+Bv^*=\Lambda v$, where $\Lambda$  is the diagonal matrix with diagonal elements $\Lambda_k\geq 0$.
Thus, since $U^\dagger \eta_k U = e^{-i \Lambda_k \Delta l}\eta_k$, the time evolved $\tilde c_n = U^\dagger c_n U $ reads
\begin{equation}
\tilde c_n = \sum_m \tilde u_{n m} c_m + \tilde v_{n m} c_m^\dagger
\end{equation}
with $\tilde u = u D u^\dagger+v D^* v^\dagger$ and $\tilde v = u D v^T+v D^* u^T$ where $D$ is the diagonal matrix with elements $D_{nn}= e^{-i \Lambda_n \Delta l} $.  Of course $\tilde E = \sum \abs{\omega_n} \tilde c^\dagger_n \tilde c_n$ is diagonal in terms of the fermions $\tilde c$. The vacuum state of the time-evolved fermion operators is $\ket{\tilde 0}$ defined by $\tilde c_n \ket{\tilde 0}=0$. We have the relation between the two vacuum states~\cite{chung01}
\begin{equation}
\ket{0} = K e^{\frac{1}{2}\sum G_{m n} \tilde c^\dagger_m\tilde c^\dagger_n}\ket{\tilde 0}
\end{equation}
with $G$ skew-symmetric defined by $\tilde u G + \tilde v =0$. Thus
\begin{equation}
e^{i u \tilde E} \ket{0} = K e^{\frac{1}{2}\sum \tilde G_{m n} \tilde c^\dagger_m\tilde c^\dagger_n}\ket{\tilde 0}
\end{equation}
with $\tilde G = \tilde D G \tilde D$ where $\tilde D$ is the diagonal matrix with elements $\tilde D_{nn}= e^{i u \abs{\omega_n}} $.
In order to calculate the inner product of Eq.~\eqref{eq.chara}, we use the fermionic coherent states defined by $\tilde c_n \ket{\xi} = \xi_n \ket{\xi}$ where $\xi_n$ are Grassmann variables~\cite{negele98}. We obtain the Gaussian integral
\begin{equation}
\bra{0} e^{i u \tilde E} \ket{0} = \abs{K}^2 \int d\xi d\xi^* e^{-\frac{1}{2}(\xi,\xi^*) \Gamma (\xi,\xi^*)^T}
\end{equation}
where
\begin{equation}
\Gamma = \left(
           \begin{array}{cc}
             G^* & -1 \\
             1 & -\tilde G \\
           \end{array}
         \right)
\end{equation}
which can be easily calculated, such that
\begin{equation}
\chi(u) = \sqrt{\frac{\det \Gamma(u)}{\det \Gamma(0)}}
\end{equation}
with $\det \Gamma(u) = \det(1+ G^\dagger \tilde G)$.
On the other hand, the characteristic function of the number of particles is similarly obtained by considering $\tilde D_{nn} = e^{iu}$.

\begin{figure}
[h!]
\includegraphics[width=0.48\columnwidth]{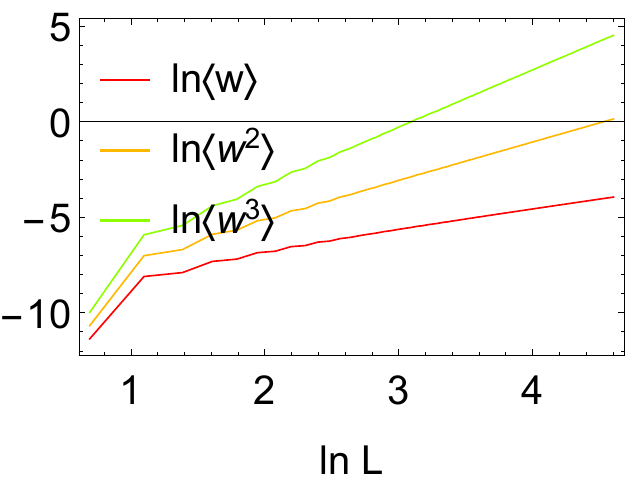} \includegraphics[width=0.48\columnwidth]{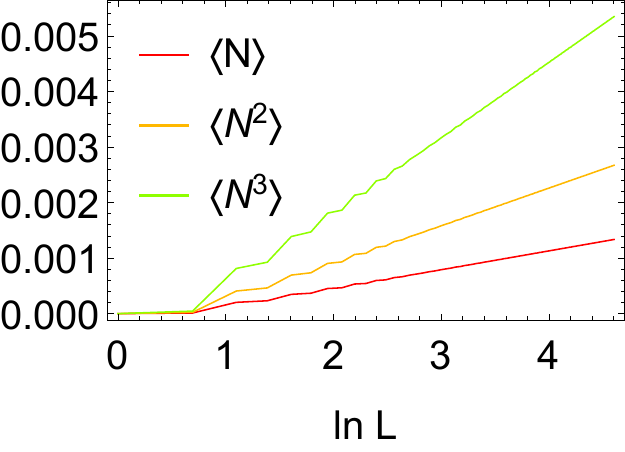}\\
\includegraphics[width=0.48\columnwidth]{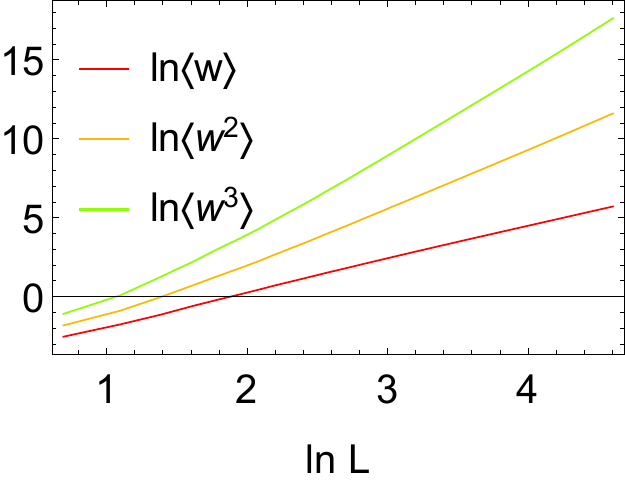} \includegraphics[width=0.48\columnwidth]{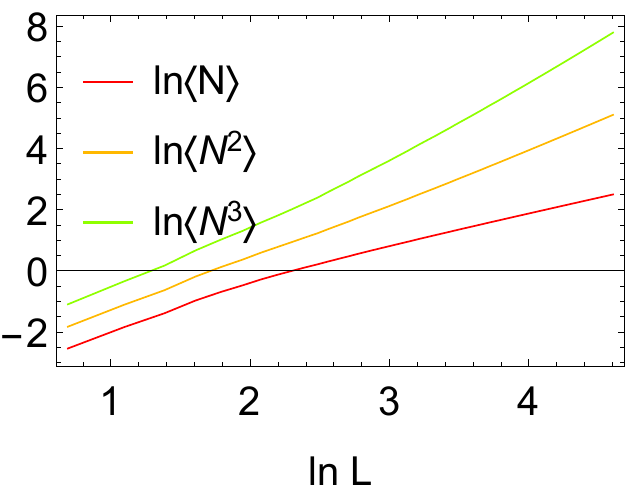}
\caption{ The plots of the moments in function of the cut-off $L$. The energy scale is in unit of $\pi v/l(t)$. We put $\Delta l = \ln 2$, $\alpha/v=0.1$ (top panels) and $\alpha/v=2$ (bottom panels).
}
\label{fig:plot1}
\end{figure}

\begin{figure}
[h!]
\includegraphics[width=0.7\columnwidth]{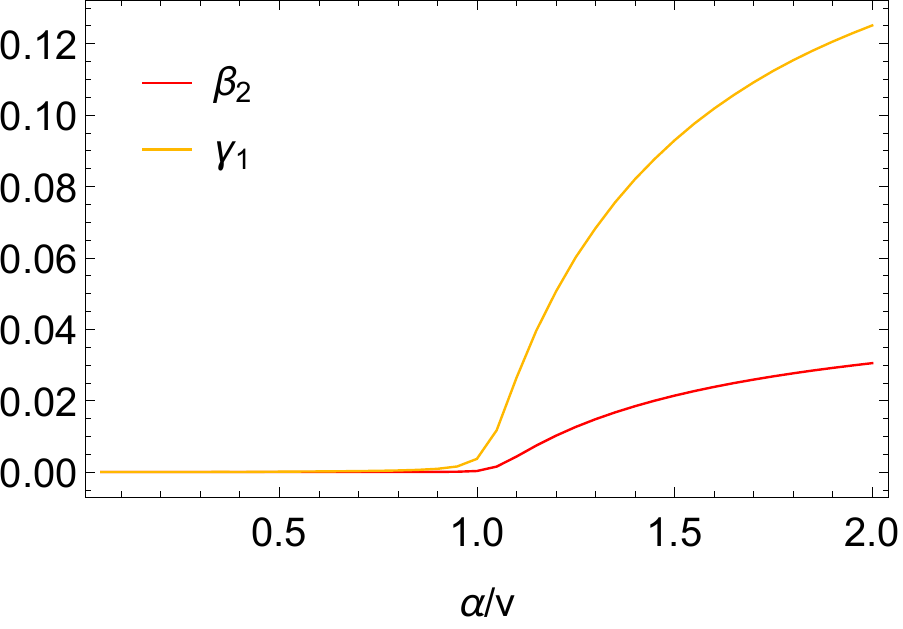}
\caption{ The plots of the coefficients $\beta_2$ and $\gamma_1$ in function of $\alpha/v$. We put $\Delta l = \ln 2$ and $\beta_2$ is in unit of $\pi v/l(t)$
}
\label{fig:plot}
\end{figure}

We calculate numerically the moments $\langle w^n \rangle$ and $\langle N^n \rangle$ by employing the respective characteristic functions and we find that the moments diverge as the number of modes $\psi_n$ goes to infinity.
By truncate the space to the $2 L$ modes $(\psi_{-L},\psi_{-L+1},\cdots,\psi_{L-1})$, we find two different behaviors depending on the value of the speed $\alpha/v$ (see Fig.~\ref{fig:plot1}).
For small $\alpha/v$ we have $\langle w^n \rangle \sim L^n$ and $\langle N^n \rangle\sim \ln L$ as $L\to \infty$.
On the other hand, for large $\alpha/v$ we get $\langle w^n \rangle \sim L^{2n}$ and $\langle N^n \rangle\sim L^n$.
We find numerical values of the parameters $\beta_\alpha,\gamma_\alpha$ that make the functions $f_w(L)=\beta_0 + \beta_1 L + \beta_2 L^2$ and $f_N(L)=\gamma_0 + \gamma_1 L + \gamma_l \ln L$ give a best fit to $\langle w \rangle$ and $\langle N \rangle$ as a function of $L$ (see Fig.~\ref{fig:plot}).
It results that $\gamma_1$ and $\beta_2$ are practically zero for small $\alpha/v$ and suddenly increase as $\alpha/v$ reaches a value $\alpha/v\approx 1$. Thus, there are two dynamical regimes: an adiabatic regime, for slow driving, where the work and the number of particles generated increase linearly and logarithmically with $L$, and a diabatic one, for fast driving, where the work and the number of particles generated increase quadratically and linearly with $L$. Of course, if $v$ is the speed of light we need a superluminal speed $\alpha$ to observe the diabatic regime.
We note that such behavior does not change for an expansion or a compression of the box. In particular, we find that the number moments $\langle N^n \rangle$ are invariant with respect to the transformation $(\alpha,\Delta l)\to - (\alpha, \Delta l)$, differently from the work moments which can change noticeably.

\section{\label{sec.conclusion} Conclusion}
In summary, we have investigated the dynamical Casimir effect in a two-dimensional massless fermionic field by adopting a canonical approach. 
We focus on the case of a linear driving of a boundary, showing how to solve the dynamics and study the statistics of work and number of particles produced from the vacuum.
By performing a numerical analysis by truncating the number of modes, we find that there are two different dynamical regimes. In both regimes, work and number of fermions generated are unbounded as the number of modes goes to infinity. In the end, we note that, for a finite cut-off $L$, for a slow driving, e.g. $\alpha/v \ll 1$, the number of particles generated can be very small, $\langle N\rangle \approx 0.03 (\alpha/v)^2\ln L$ for $\Delta l = \ln 2$, but the work produced can be relatively big, $\langle w\rangle\approx 0.02 (\alpha/v)^2 L \pi v/l(t)$, which can be observed as heat given to the  environment in a thermalization process at zero temperature~\cite{plastina14}.

\end{document}